\documentclass[reprint, superscriptaddress, twocolumn, amsmath, amssymb, aps, prb]{revtex4-2}
\usepackage{graphicx}
\usepackage{dcolumn}
\usepackage{bm}
\usepackage{placeins}
\usepackage{appendix}
\usepackage{upgreek}

\usepackage{verbatim}
\usepackage[usenames,dvipsnames]{xcolor}
 \usepackage{amsmath}
 \usepackage{amsfonts}
 \usepackage{amssymb}
 \usepackage[colorlinks=true,citecolor=blue,linkcolor=red]{hyperref}

 \usepackage{bbold}     
 \usepackage[makeroom]{cancel}  
 \usepackage{multirow}    
 \usepackage[normalem]{ulem}        
 \usepackage{array}
 \usepackage{pdfpages}
\makeatletter
 \AtBeginDocument{\let\LS@rot\@undefined}
 \makeatother

\begin{document}


\title{Ballistic PbTe Nanowire Devices}

\author{Yuhao Wang}
 \email{equal contribution}
\affiliation{State Key Laboratory of Low Dimensional Quantum Physics, Department of Physics, Tsinghua University, Beijing 100084, China}

\author{Fangting Chen}
\email{equal contribution}
\affiliation{State Key Laboratory of Low Dimensional Quantum Physics, Department of Physics, Tsinghua University, Beijing 100084, China}

\author{Wenyu Song}
\email{equal contribution}
\affiliation{State Key Laboratory of Low Dimensional Quantum Physics, Department of Physics, Tsinghua University, Beijing 100084, China}

\author{Zuhan Geng}
\affiliation{State Key Laboratory of Low Dimensional Quantum Physics, Department of Physics, Tsinghua University, Beijing 100084, China}

\author{Zehao Yu}
\affiliation{State Key Laboratory of Low Dimensional Quantum Physics, Department of Physics, Tsinghua University, Beijing 100084, China}

\author{Lining Yang}
\affiliation{State Key Laboratory of Low Dimensional Quantum Physics, Department of Physics, Tsinghua University, Beijing 100084, China}

\author{Yichun Gao}
\affiliation{State Key Laboratory of Low Dimensional Quantum Physics, Department of Physics, Tsinghua University, Beijing 100084, China}

\author{Ruidong Li}
\affiliation{State Key Laboratory of Low Dimensional Quantum Physics, Department of Physics, Tsinghua University, Beijing 100084, China}

\author{Shuai Yang}
\affiliation{State Key Laboratory of Low Dimensional Quantum Physics, Department of Physics, Tsinghua University, Beijing 100084, China}

\author{Wentao Miao}
\affiliation{State Key Laboratory of Low Dimensional Quantum Physics, Department of Physics, Tsinghua University, Beijing 100084, China}

\author{Wei Xu}
\affiliation{State Key Laboratory of Low Dimensional Quantum Physics, Department of Physics, Tsinghua University, Beijing 100084, China}

\author{Zhaoyu Wang}
\affiliation{State Key Laboratory of Low Dimensional Quantum Physics, Department of Physics, Tsinghua University, Beijing 100084, China}

\author{Zezhou Xia}
\affiliation{State Key Laboratory of Low Dimensional Quantum Physics, Department of Physics, Tsinghua University, Beijing 100084, China}

\author{Huading Song}
\affiliation{Beijing Academy of Quantum Information Sciences, Beijing 100193, China}

\author{Xiao Feng}
\affiliation{State Key Laboratory of Low Dimensional Quantum Physics, Department of Physics, Tsinghua University, Beijing 100084, China}
\affiliation{Beijing Academy of Quantum Information Sciences, Beijing 100193, China}
\affiliation{Frontier Science Center for Quantum Information, Beijing 100084, China}
\affiliation{Hefei National Laboratory, Hefei 230088, China}

\author{Yunyi Zang}
\affiliation{Beijing Academy of Quantum Information Sciences, Beijing 100193, China}
\affiliation{Hefei National Laboratory, Hefei 230088, China}

\author{Lin Li}
\affiliation{Beijing Academy of Quantum Information Sciences, Beijing 100193, China}

\author{Runan Shang}
\affiliation{Beijing Academy of Quantum Information Sciences, Beijing 100193, China}
\affiliation{Hefei National Laboratory, Hefei 230088, China}

\author{Qi-Kun Xue}
\affiliation{State Key Laboratory of Low Dimensional Quantum Physics, Department of Physics, Tsinghua University, Beijing 100084, China}
\affiliation{Beijing Academy of Quantum Information Sciences, Beijing 100193, China}
\affiliation{Frontier Science Center for Quantum Information, Beijing 100084, China}
\affiliation{Hefei National Laboratory, Hefei 230088, China}
\affiliation{Southern University of Science and Technology, Shenzhen 518055, China}

\author{Ke He}
\email{kehe@tsinghua.edu.cn}
\affiliation{State Key Laboratory of Low Dimensional Quantum Physics, Department of Physics, Tsinghua University, Beijing 100084, China}
\affiliation{Beijing Academy of Quantum Information Sciences, Beijing 100193, China}
\affiliation{Frontier Science Center for Quantum Information, Beijing 100084, China}
\affiliation{Hefei National Laboratory, Hefei 230088, China}

\author{Hao Zhang}
\email{hzquantum@mail.tsinghua.edu.cn}
\affiliation{State Key Laboratory of Low Dimensional Quantum Physics, Department of Physics, Tsinghua University, Beijing 100084, China}
\affiliation{Beijing Academy of Quantum Information Sciences, Beijing 100193, China}
\affiliation{Frontier Science Center for Quantum Information, Beijing 100084, China}


\begin{abstract}

Disorder is the primary obstacle in current Majorana nanowire experiments. Reducing disorder or achieving ballistic transport is thus of paramount importance. In clean and ballistic nanowire devices, quantized conductance is expected with plateau quality serving as a benchmark for disorder assessment. Here, we introduce ballistic PbTe nanowire devices grown using the selective-area-growth (SAG) technique. Quantized conductance plateaus in units of $2e^2/h$ are observed at zero magnetic field. This observation represents an advancement in diminishing disorder within SAG nanowires, as none of the previously studied SAG nanowires (InSb or InAs) exhibit zero-field ballistic transport. Notably, the plateau values indicate that the ubiquitous valley degeneracy in PbTe is lifted in nanowire devices. This degeneracy lifting addresses an additional concern in the pursuit of Majorana realization. Moreover, these ballistic PbTe nanowires may enable the search for clean signatures of the spin-orbit helical gap in future devices.  

\end{abstract}

\maketitle  

Electron transport in a semiconductor nanowire is highly sensitive to disorder, owing to its large surface-to-volume ratio and the lack of screening. Consequently, transport in such systems is commonly diffusive. Reducing disorder in nanowires is a challenging task, but urgently needed \cite{GoodBadUgly} for its potential applications in quantum devices, e.g. the realization of Majorana zero modes. A clean semiconductor nanowire coupled to a superconductor has been predicted to host Majorana zero modes  \cite{Lutchyn2010, Oreg2010}. Despite tremendous efforts being devoted to reducing disorder in InAs and InSb nanowire devices \cite{Chang2015, Krogstrup2015, Kammhuber2016, Gul2017, Zhang2017Ballistic}, all the Majorana nanowire experiments \cite{Mourik, Deng2016, Gul2018, Song2022, NextSteps, Prada2020,cao2022recent} so far still suffer from disorder. Take, for instance, the four recent studies: the observation of quantized zero-bias peaks in a thin InAs-Al nanowire \cite{WangZhaoyu}, simulating Kitaev chain using two quantum dots coupled to a superconductor \cite{Delft_Kitaev}, the topological-gap-protocol devices based on three-terminal InAs-Al \cite{MS_2023}, and the zero-bias peaks in phase-tunable planar Josephson junctions \cite{PRB_Marcus_2023_Planar}. All these studies have used a gate-tunable quantum point contact (QPC) as the probe, yet none of them can reveal quantized plateaus for their normal state conductance. The quantized conductance is a characteristic signature of ballistic QPCs \cite{Wharam_1988, vanWees_1988} and signifies a minimal degree of disorder. The absence of ballistic transport implies that significant amount of disorder is present or even dominating in those III-V semiconductor experiments. Moreover, extensive theory studies have also strongly suggested that disorder is the current roadblock in Majorana research \cite{Patrick_Lee_disorder_2012, Pikulin_2012, Loss2013ZBP,GoodBadUgly, DasSarma2021Disorder, Tudor2021Disorder}.    

Here, we explore PbTe nanowires, an IV-VI semiconductor, as an alternative platform. We demonstrate ballistic transport in these wires that may circumvent the challenges posed by disorder. PbTe nanowires have recently attracted much interest due to their large dielectric constant ($\sim$1350), which can screen charge disorder \cite{CaoZhanPbTe}. Experimentally, these nanowires and networks have been successfully grown using the selective-area-growth (SAG) technique \cite{Jiangyuying, Erik_PbTe_SAG}. Basic transport characterizations, including phase coherent transport \cite{PbTe_AB}, quantum dots  \cite{Fabrizio_PbTe}, the superconducting proximity effect  \cite{Zitong}, and the hard gap \cite{Yichun}, have been demonstrated. While quantized conductance plateaus have been observed \cite{Wenyu}, they require a large magnetic field ($B \sim$ 6 T). The zero-field transport remains diffusive. Recently, we have reduced the device disorder. In this study, we report the observation of quantized conductance in PbTe nanowires at zero magnetic field, indicative of ballistic transport. Previously, zero-field quantization has been realized in III-V nanowires \cite{abay2013quantized, Kammhuber2016, InAs_Gooth, Silvano_ballistic}, mainly grown using the vapor-liquid-solid technique. Those wires are, however, difficult to scale up for future applications. Notably, none of the previous SAG nanowires could reveal zero-field ballistic transport \cite{2018_PRL_SAG, Friedl, Palmstrom_SAG, Pavel_SAG_2019, Roy}, making our observations a step forward in enhancing the quality of SAG wires.  Another potential application, closely aligned with Majorana research, involves the helical gap induced by spin-orbit interaction \cite{2004_PRB_Helical_theory}. Despite years of searching, the re-entrance behavior observed in earlier experiments \cite{GG_helical, 2017_Jakob_Helical, Heedt_Helical} may possibly be disorder-induced \cite{PRB_2014_helical_gap}. Our devices could serve as a clean platform to identify signatures of helical gaps.

\begin{figure}[htb]
\includegraphics[width=\columnwidth]{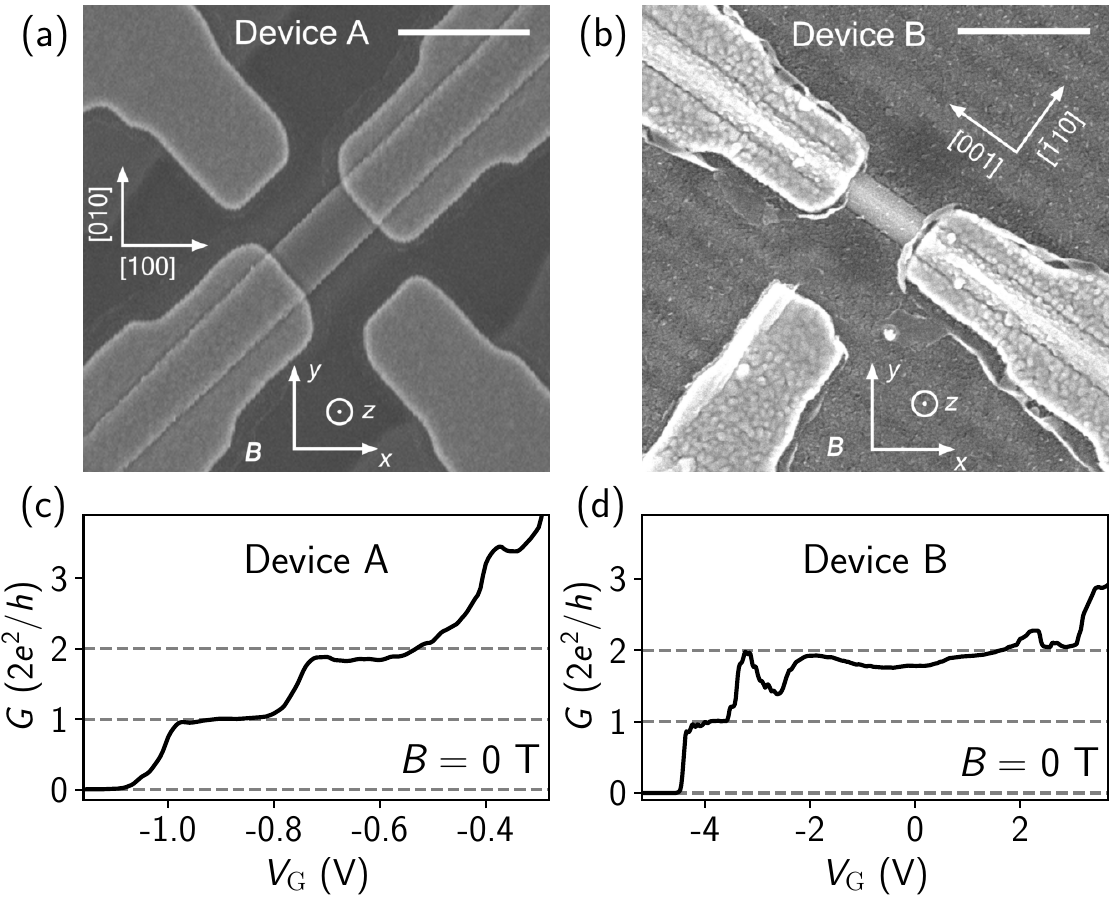}
\centering
\caption{Device basics. (a-b) SEMs of devices A and B, respectively. Crystal directions are labeled. Scale bar, 500 nm. (c-d) $G$ vs $V_{\text{G}}$ in devices A and B. (c) is a line cut extracted from Fig. 3(a). $V$ = 0 mV. $B$ = 0 T.  }
\label{fig1}
\end{figure}

Figures 1(a) and 1(b) illustrate scanning electron micrographs (SEMs) of two devices, denoted as A and B. The PbTe nanowires were selectively grown on a CdTe substrate \cite{Jiangyuying}. A thin insulating layer of Pb$_{1-x}$Eu$_x$Te was grown prior to the PbTe growth, which could suppress disorder \cite{2021_Boris}. The value of $x$ is estimated to be 0.08 for device A and 0.07 for device B. The substrate of device A is CdTe(001) while for device B it is CdTe(110). After the PbTe growth, a thin layer of Pb$_{1-x}$Eu$_x$Te (CdTe) was capped for device A (B) to prevent oxidization. Ohmic contacts and side gates were deposited via the evaporation of Ti/Au. The thickness of Ti is $\sim$ 10 nm, and for Au it is 55 (75) nm for device A (B).  The two side gates of device A were interconnected during the measurement.

The linear conductance as a function of the gate voltage ($V_{\text{G}}$) for both devices exhibits quantized plateaus, as shown in Figs. 1(c-d).  $B$ = 0 T. These measurements were performed in a dilution refrigerator operating at its base temperature ($\sim$ 50 mK). Standard two-terminal circuit was used to obtain the device differential conductance, $G \equiv dI/dV$. The bias voltage $V$ between the two contacts was fixed to zero. The line-resistance, mainly contributed by the fridge filters, was subtracted. Contact resistances of 900 $\Omega$ for device A and 500 $\Omega$ for device B were estimated (and subtracted) based on the plateau values.

\begin{figure}[ht]
\includegraphics[width=\columnwidth]{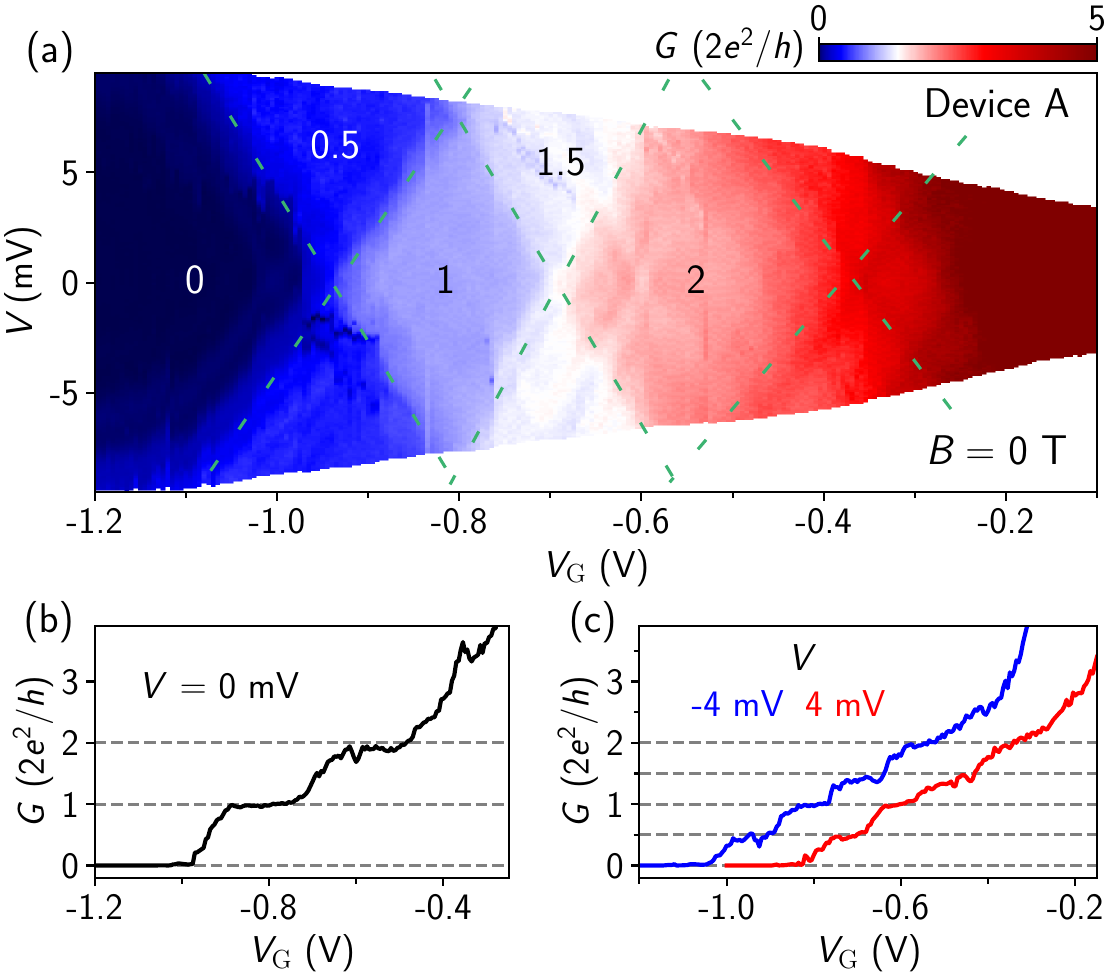}
\centering
\caption{Zero-field quantization in device A. (a) $G$ vs $V_{\mathrm{G}}$ and $V$. $B$ = 0 T. Dashed lines are guide for eyes to mark the diamond edges. (b) Zero-bias line cut from (a). (c) Line cuts at $V=$ 4 mV and -4 mV from (a). The red curve was offset by 0.2 V horizontally for clarity.  }
\label{fig2}
\end{figure}

The zero-field plateaus are in steps of $2e^2/h$, indicating the lifting of valley degeneracy in PbTe nanowires on both (001) and (110) substrates. The pre-factor of 2 originates from spin degeneracy. Earlier numerical calculations had suggested a two-fold valley degeneracy along these crystal orientations \cite{CaoZhanPbTe}. Our recent work has indeed confirmed this two-fold degeneracy by observing $2e^2/h$ plateaus at high magnetic fields along the same orientation as device A \cite{Wenyu}. In the current study, we have reduced the wire thickness to $\sim$ 40 nm for device A (as opposed to $\sim$ 100 nm in our previous work \cite{Wenyu}). We thus attribute the lifting of valley degeneracy in device A to a stronger confinement. For device B, the nanowire top surface is triangle-like. Note that the valley degeneracy could be lifted by device asymmetry due to the capping/substrate layer, the side mask or a side gate.

Conductance oscillations are present and superimposed on the plateaus of device B in Fig. 1(d). The oscillation amplitudes are large for the second plateau and small for the first one. We ascribe the small oscillations on the first plateau to be Fabry-P\'erot-induced, a phenomenon typical in QPCs with sharp confinement potentials \cite{1994_PRB_QPC_FB}. The large modulations on the second plateau likely arise from a quasi-bound state, which will be discussed later.

\begin{figure*}[t]
\includegraphics[width=0.9\textwidth]{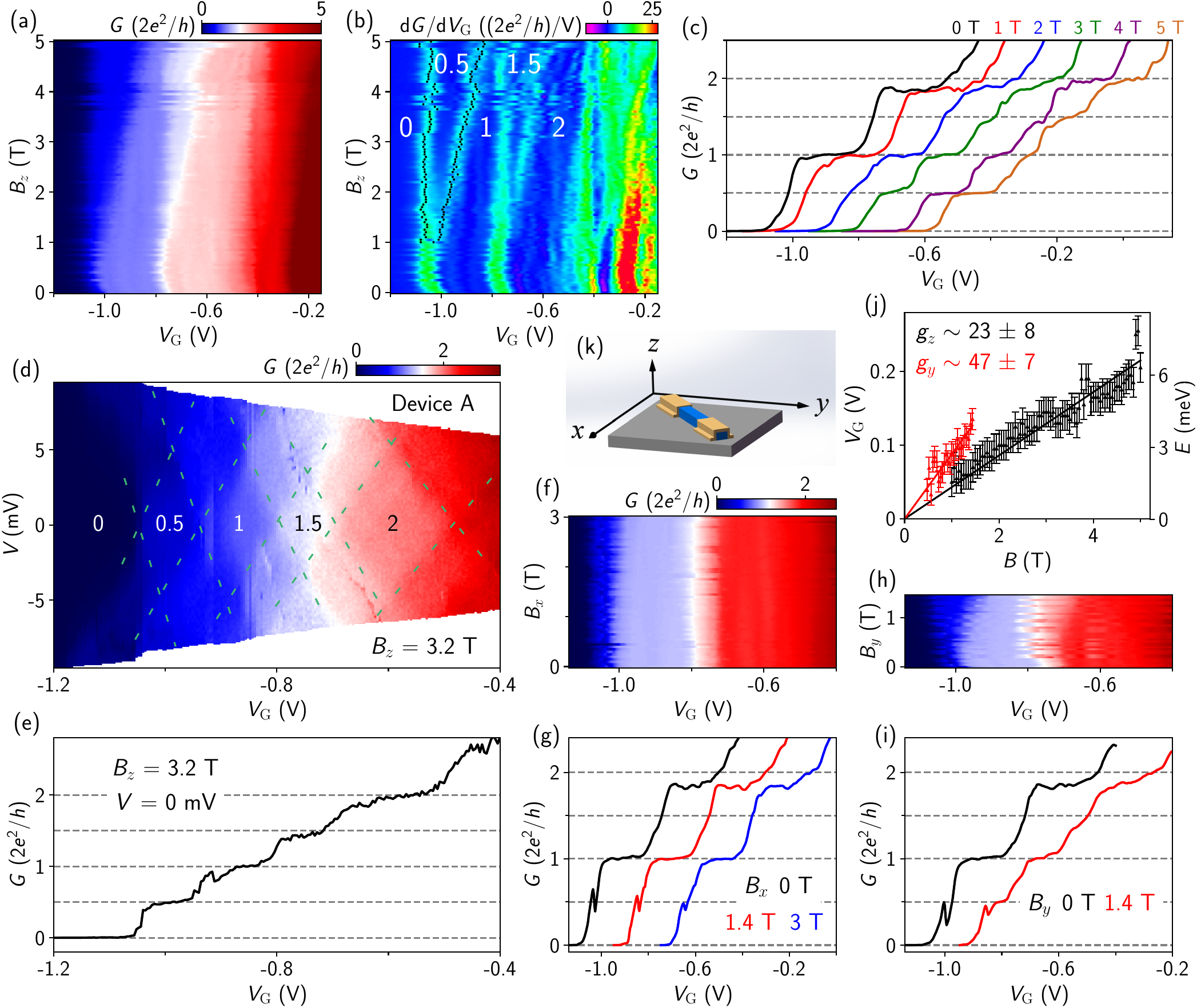}
\centering
\caption{$B$ dependence in device A. (a) $G$ vs $V_{\text{G}}$ and $B_{z}$ (direction, out-of-plane). $V$ = 0  mV. (b) Transconductance plot of (a) with  minor smoothing over $V_{\text{G}}$. Labeled numbers are the plateau values in units of $2e^2/h$. The black dots mark the positions of $G = 0.25$ and 0.75 (in units of $2e^2/h$), agreeing well with the transconductance peaks. (c) Line cuts of (a) with a horizontal offset of 0.1 V between neighboring curves. (d) $G$ vs $V$ and $V_{\text{G}}$ at $B_z$ = 3.2 T. The dashed lines mark the plateau diamonds. (e) Zero-bias line cut of (d). (f) $B$ scan along $x$ axis. (g) Line cuts from (f) with a horizontal offset of 0.2 V between neighboring curves. (h) $B$ scan along $y$ axis. (i) Line cuts from (h) with a horizontal offset of 0.2 V. (j) Width of the $e^2/h$ plateau as a function of $B_z$ (black) and $B_y$ (red). Error bars are estimated based on the widths of the transconductance peaks. The solid lines are linear fits.  (k) Sketch of the coordinate axes. }
\label{fig3}
\end{figure*}

\begin{figure*}[t]
\includegraphics[width=\textwidth]{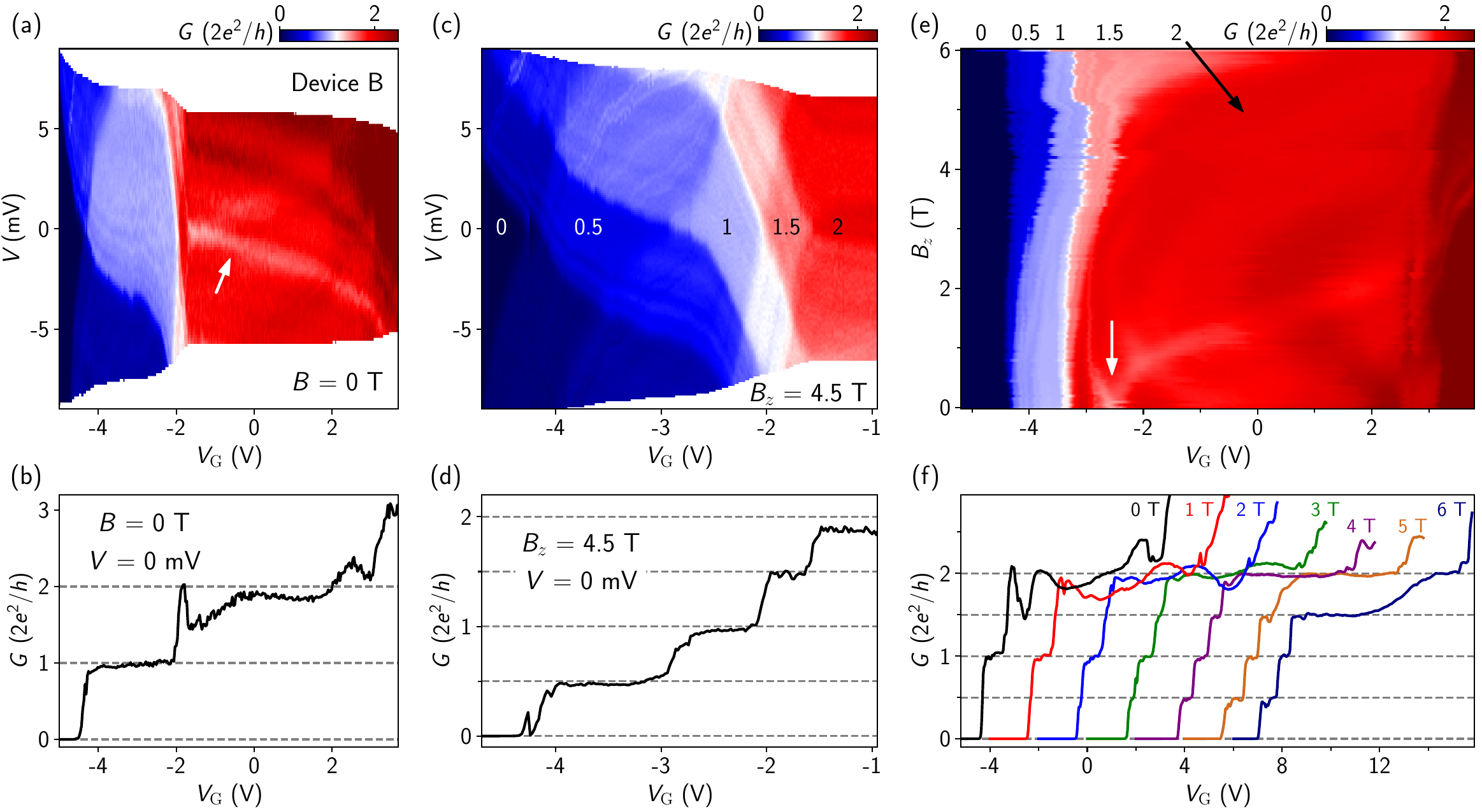}
\centering
\caption{Ballistic transport in device B. (a) $G$ vs $V_{\text{G}}$ and $V$ at $B$ = 0 T. (b) Zero-bias line cut of (a). (c) $G$ vs $V_{\text{G}}$ and $V$ at $B_z$ = 4.5 T. (d) Zero-bias line cut of (c). (e) $B_z$ scan of the plateaus. $V$ = 0 mV. The numbers labeled on the top denote the plateau values (in units of $2e^2/h$).  (f) Several line cuts from (e) with a horizontal offset of 2 V between neighboring curves. }
\label{fig4}
\end{figure*}

To reveal the sizes of subband spacing, Fig. 2(a) shows the 2D conductance map, $G$ vs $V$ and $V_{\text{G}}$. For the waterfall and transconductance plots of Fig. 2(a), see Fig. S1 in the Supplemental Material. The zero-bias plateaus are now revealed as diamond shapes within which the conductance remains nearly constant. The diamond values are labeled in units of $2e^2/h$. The diamond boundaries (dashed lines) correspond to the alignment of the quasi-fermi levels in the source/drain contacts with each subband. The sizes of the diamonds thus provide a direct measurement of the energy spacing between subbands.  We estimate the spacing between the second subband ($E_2$) and the first one ($E_1$) to be $E_2-E_1 \sim$ 8.3 meV. Likewise, the spacing $E_3-E_2 \sim$ 9.6 meV can also be estimated,  with $E_3$ denoting the third subband. Moreover, the half plateaus at finite bias correspond to situations in which the quasi-fermi levels of the source and drain contacts are positioned within different subbands. Line cuts at both zero and finite bias for integer and half plateaus are shown in Figs. 2(b-c).

Next we study the $B$ dependence of these plateaus. Figure 3(a) shows the $B$ scan along the $z$ axis, which is perpendicular to the nanowire (out-of-plane). The coordinate axes are drawn in the middle panel of Fig. 3 (also labeled in Fig. 1(a)). As $B$ is increased, the $e^2/h$ and $3e^2/h$ plateaus gradually emerge and expand from the $2e^2/h$ and $4e^2/h$ plateaus. These half plateaus (in units of $2e^2/h$) arise due to the Zeeman splitting of subbands. The transconductance, $dG/dV_{\text{G}}$ in Fig. 3(b), can better illustrate this evolution, see also Fig. 3(c) for line cuts.

Figure 3(d) presents the 2D scan of $G$ vs $V$ and $V_{\text{G}}$ with $B_z$ being fixed at 3.2 T. The dashed lines highlight the plateau diamonds, with labeled values provided in units of $2e^2/h$. For the waterfall and transconductance plots, see Fig. S1. The zero-bias line cut in Fig. 3(e) illustrates the plateaus, accompanied by charge jumps, e.g. near $V_{\text{G}}$ = -0.9 V. The charge jump occurred swiftly (within a single pixel of the color map), leading to a sharp conductance drop. Obviously, this re-entrance behavior is not the signature of a  helical gap. 

Figures 3(f-i) are the $B$ scans along two other directions. The $B$ range is limited by hardware. The small peaks below the first plateau are charge jumps. Notably, the $e^2/h$ plateau is already well developed for $B_y$ = 1.4 T (the red curve in Fig. 3(i)). As a contrast, this half plateau is absent for $B_x$ at the same field (the red curve in Fig. 3(g)), and also barely visible for $B_z$ at an even higher field of 2 T (the blue curve in Fig. 3(c)). This anisotropic behavior indicates variations in the Land\'{e} $g$-factor for different $B$ directions. 

In Fig. 3(j) we extract the width of the $e^2/h$ plateau as a function of $B$ along the $z$ and $y$ axes. The width is estimated based on the spacing between the black dots (peaks in the transcodnuctance) in Figs. 3(b) and S1(f). The $B_x$ scan does not reveal the $e^2/h$ plateau at the highest applied field (3 T), thus not presented. The plateau width in $V_{\text{G}}$ is converted to energy based on the lever arm extracted from Fig. 2(a). This energy corresponds to the Zeeman splitting $|E_{1\downarrow}-E_{1\uparrow}|=g\mu_{\text{B}}B$. $\mu_{\text{B}}$ is the Bohr magneton and the arrows denote spins. The slopes of the linear fits give a $g$-factor of 23 $\pm$ 8 for $B_z$ and 47 $\pm$ 7 for $B_y$. Note that the presence of a charge jump near $e^2/h$ plateau may lead to a slight overestimate of the corresponding $g$-factor.  For the analysis of other plateaus and the corresponding $g$-factors, see Fig. S1. The $g$-factor anisotropy in nanowires is widely present \cite{Fabrizio_PbTe} and depends on various factors such as confinement and crystal direction \cite{CaoZhanPbTe, g-factor_reduction}.

Figure 4 shows a second device (device B) that exhibits ballistic transport. The first plateau appears as a twisted diamond in Fig. 4(a). This twist implies a non-linear crosstalk between bias and gate voltages. The origin of this crosstalk is currently unknown. The size of the diamond suggests a subband spacing $E_2-E_1 \sim$ 8 meV. The second plateau in Fig. 4(b) is severely modulated by a quasi-bound state, a resonant level formed by weak scattering within the device. Its energy is gate tunable, see the white arrow in Fig. 4(a). These resonant-induced oscillations can mimic the re-entrance behavior characteristic of a helical gap \cite{PRB_2014_helical_gap}. 

At $B_z$ = 4.5 T, the first four spin-resolved plateaus, in steps of $e^2/h$, are clearly revealed in Figs. 4(c-d). The diamonds of the 0.5 and 1 plateaus (in units of $2e^2/h$) also exhibit a twisted shape. The $4e^2/h$ plateau in Fig. 4(d) is slightly lower than the quantized value, suggesting a non-unity transmission for that particular subband. We estimate a $g$-factor of $\sim$ 15 based on the size of the 0.5 plateau ($\sim$ 4 meV). Fabry-P\'erot-like oscillations are visible within the plateau diamonds in Fig. 4(c), suggesting a sharp QPC constriction potential \cite{1994_PRB_QPC_FB}.

\begin{figure*}[htb]
\includegraphics[width=\textwidth]{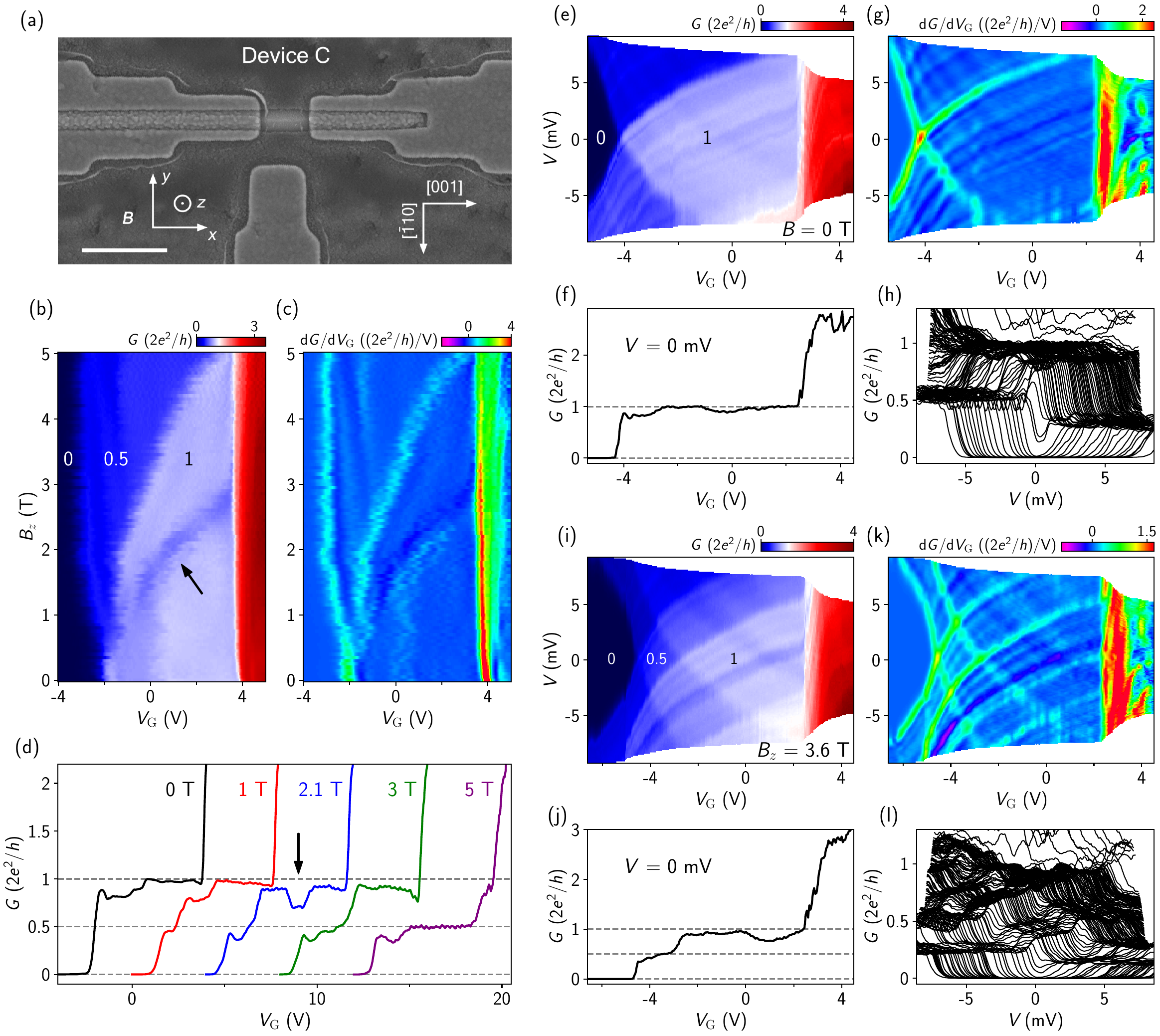}
\centering
\caption{Ballistic transport in device C. (a) SEM. Scale bar, 500 nm. (b) $B_z$ scan. $V$ = 0 mV. The labeled numbers are the plateau values in units of $2e^2/h$.  (c) Transconductance plot of (b) with minor smoothing. (d) Line cuts from (b) at different $B_z$'s (labeled) with a horizontal offset of 4 V between neighboring curves.  (e) $G$ vs $V$ and $V_{\text{G}}$ at $B$ = 0 T. (f) Zero-bias line cut of (e). (g) Transconductance of (e) with minor smoothing. (h) Waterfall plot of (e). (i) $G$ vs $V$ and $V_{\text{G}}$ at $B_z$ = 3.6 T. (j) Zero-bias line cut of (i). (k) Transconductance of (i) with minor smoothing. (l) Waterfall plot of (i).  } 
\label{fig5}
\end{figure*}

In Fig. 4(e) we show the $B_z$ evolution of the plateaus with line cuts shown in Fig. 4(f). The resonance level on the second plateau in Fig. 4(e) quickly splits upon increasing $B$ (see the white arrow), possibly due to Zeeman splitting. As a result, the resonant-level-induced oscillations are suppressed at higher fields, see Fig. 4(f).  For the waterfall plots of Figs. 4(a) and 4(c), the transconductance of Figs. 4(a), 4(c) and 4(e), and the $g$-factor estimations, see Fig. S2 in the Supplemental Material.

We further present a third ballistic device (device C), see the SEM in Fig. 5(a). The substrate is CdTe(110). Figures 5(b-d) illustrate the $B_z$ scan, the transconductance plot and line cuts. A contact resistance of 1.1 k$\Omega$ was subtracted. The $2e^2/h$ plateau is clearly revealed at zero field. The absence of the $4e^2/h$ plateau indicates valley or orbital degeneracy of the second subband. As $B_z$ is increased, the $e^2/h$ plateau emerges and expands due to Zeeman splitting.

Notably, an additional dip feature (see the arrow in Figs. 5(b) and 5(d)) also develops on the $2e^2/h$ plateau. The blue line cut in Fig. 5(d) depicts this drop, a re-entrance of conductance, on the $2e^2/h$ plateau. This re-entrance behavior, obviously not a charge jump, is reminiscent of a helical gap. For a continuous evolution of the dip, see all the line cuts ($B_z$ from 0 T to 4 T) of this feature in Fig. S3. The $B_y$ and $B_x$ scans, also shown in Fig. S3, do not reveal this re-entrance signature, possibly due to the $g$-factor anisotropy. The dip does not reach $e^2/h$, while a conductance dip of $e^2/h$ is expected in the ideal situation of a helical gap. Furthermore, the dip bends towards higher energy (more positive $V_{\text{G}}$) for higher $B_z$, where in the simplest helical gap model, the re-entrance dip should gradually merge toward the $e^2/h$ plateau. These deviations suggest that either the observation does not arise from a helical gap or a more sophisticated helical gap model is required. Unlike III-V semiconductors, the highly anisotropic properties of PbTe does lead to a complicated situation, e.g. $B_z$ could also induce Zeeman splitting for spins along $x$ and $y$ axis, as the $g$-factor in PbTe is a tensor \cite{CaoZhanPbTe} and may be $B$ dependent. More theory inputs and experimental studies in more devices with different wire geometries and orientations are required before one can conclude on the origin of this re-entrance observation. 

Figures 5(e-h) show the 2D map of the first plateau, the transconductance and the waterfall plots at zero field while Figs. 5(i-l) are for $B_z$ = 3.6 T. The diamond size in Fig. 5(e) suggests a subband spacing $E_2-E_1 \sim$ 8.7 meV. Fabry-P\'erot oscillations are present and superimposed on the plateau diamonds, shown in Figs. 5(e) and 5(g). The first plateau in Fig. 5(f) is revealed as clusters of curves near $2e^2/h$ in the waterfall plots in Fig. 5(h).  The clusters at high biases, e.g. $V = \pm$5 mV, are the fractional plateaus. The fractional values are not exactly half, probably due to the asymmetry of the device circuit. The lock-in excitation ($dV$) is not equally shared between the quasi-fermi levels of the source and drain contacts. The diamond size of the 0.5 plateau in Fig. 5(i) is $\sim$ 3.2 meV, indicating a $g$-factor of $\sim$ 15. Besides the three devices (A-C), Figs. S4-S6 present three additional devices that exhibit signatures of ballistic transport at zero field. 

In summary, we have observed quantized conductance plateaus in units of $2e^2/h$ at zero magnetic field in PbTe nanowires. The plateau values indicate the lifting of valley degeneracy in PbTe. At finite magnetic fields, half plateaus can be observed due to Zeeman splitting. Land\'{e} $g$-factor can be extracted and its anisotropy has been discussed. A conductance dip on the $2e^2/h$ plateau is observed in one device, reminiscent of (also with deviations from) the possible re-entrance behavior of a helical gap. These ballistic PbTe nanowire devices represent an improvement on diminishing disorder in SAG nanowires, which may advance the Majorana research and the search of spin-orbit helical gaps.

\section{Acknowledgment} 

We thank Leo Kouwenhoven for valuable discussions. This work is supported by Tsinghua University Initiative Scientific Research Program, National Natural Science Foundation of China (92065206) and the Innovation Program for Quantum Science and Technology (2021ZD0302400).

\section{Data Availability} 

Raw data and processing codes within this paper are available at   https://doi.org/10.5281/zenodo.8336920

\bibliography{mybibfile}

\newpage

\onecolumngrid

\newpage


\section*{Supplementary material}

Yuhao Wang,[1, *] Fangting Chen,[1, *] Wenyu Song,[1, *] Zuhan Geng,[1] Zehao Yu,[1] Lining Yang,[1] Yichun Gao,[1] Ruidong Li,[1] Shuai Yang,[1] Wentao Miao,[1] Wei Xu,[1] Zhaoyu Wang,[1] Zezhou Xia,[1] Huading Song,[2] Xiao Feng,[1, 2, 3, 4] Yunyi Zang,[2, 4] Lin Li,[2] Runan Shang,[2, 4] Qi-Kun Xue,[1, 2, 3, 4, 5] Ke He,[1, 2, 3, 4, †] and Hao Zhang[1, 2, 3, ‡]

[1]State Key Laboratory of Low Dimensional Quantum Physics,
Department of Physics, Tsinghua University, Beijing 100084, China
[2]Beijing Academy of Quantum Information Sciences, Beijing 100193, China
[3]Frontier Science Center for Quantum Information, Beijing 100084, China
[4]Hefei National Laboratory, Hefei 230088, China
[5]Southern University of Science and Technology, Shenzhen 518055, China

**Additional information about the measurement and data analysis**

Device growth and fabrication are similar to those reported in our previous works [1, 2]. Details of modification in growth will be presented and analyzed elsewhere. In the two-terminal measurement, a series resistance, $R_{series}$, has been subtracted from $G$. $R_{series} = R_{fridge} + R_{contact}$. $R_{fridge}$ is mainly the filter resistance and input/output resistance of the circuit. The filter resistance is ∼ 3.5 kΩ for devices B and F, and ∼ 2.3 kΩ for other devices, as they were measured in different fridges. $R_{fridge}$ was measured independently by shorting two bonding pads at the mixing chamber. $R_{contact}$ is the device contact resistance and estimated based on the plateau values. The bias voltage shared by $R_{series}$ has also been subtracted from $V$.

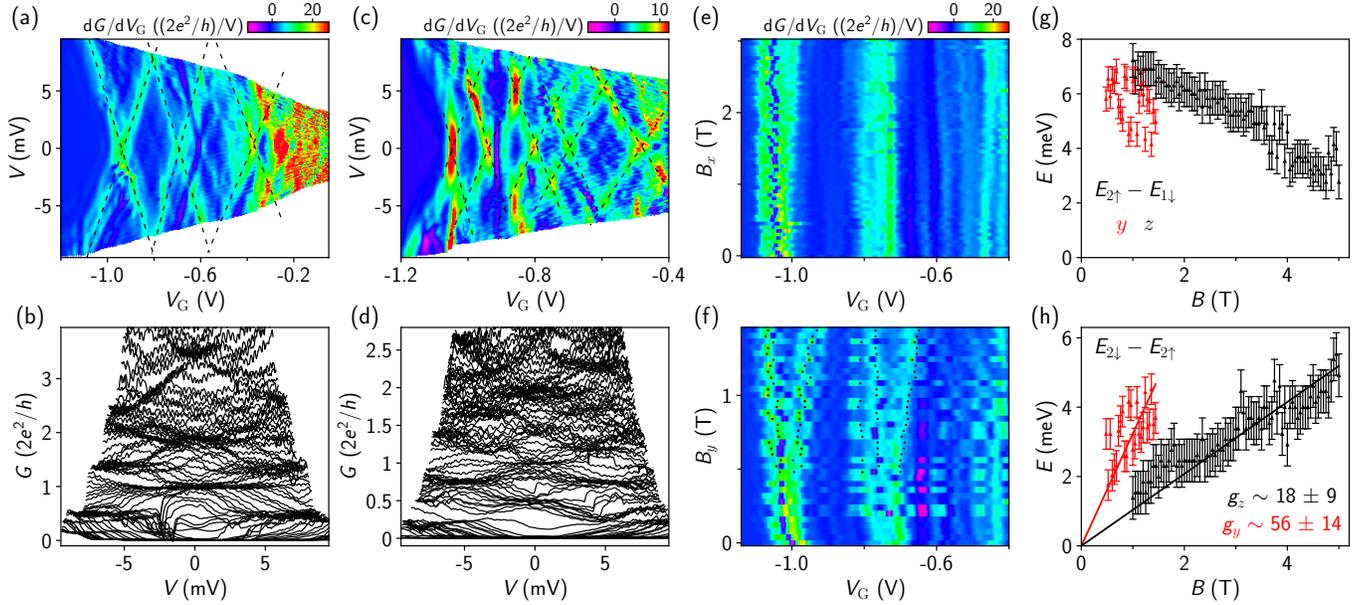

FIG. S1. Additional analysis of device A. (a) Transconductance plot of Fig. 2(a) with minor smoothing. The dashed lines highlight the plateau diamonds. (b) Waterfall plot of Fig. 2(a). Every other curve was plotted for clarity. (c) Transconductance plot of Fig. 3(d) with minor smoothing. The dashed lines highlight the plateau diamonds. (d) Waterfall plot of Fig. 3(d). Every other curve was plotted for clarity. (e-f), Transconductance plots of Figs. 3(f) and 3(h), respectively. Zeeman splitting of the first plateau is observable for $B_y$ and absent for $B_z$, indicating the g-factor anisotropy. The dots in (f) mark the positions where $G = 0.25$, $0.75$, $1.25$ and $1.75$ (in units of $2e^2/h$) in Fig. 3(h), agreeing well with the transconductance peaks. (g) Width of the $2e^2/h$ plateau in Figs. 3(a) and 3(h), i.e. $E_{2\uparrow} - E_{1\downarrow}$. (h) Width of the $3e^2/h$ plateau in Figs. 3(a) and 3(h), i.e. $E_{2\downarrow} - E_{2\uparrow}$. The energy was converted from $V_G$ based on the extracted lever arm. The linear fits suggest a g-factor of $18 \pm 9$ for $B_z$ and $56 \pm 14$ for $B_y$ for the second subband.

* equal contribution
† kehe@tsinghua.edu.cn
‡ hzquantum@mail.tsinghua.edu.cn

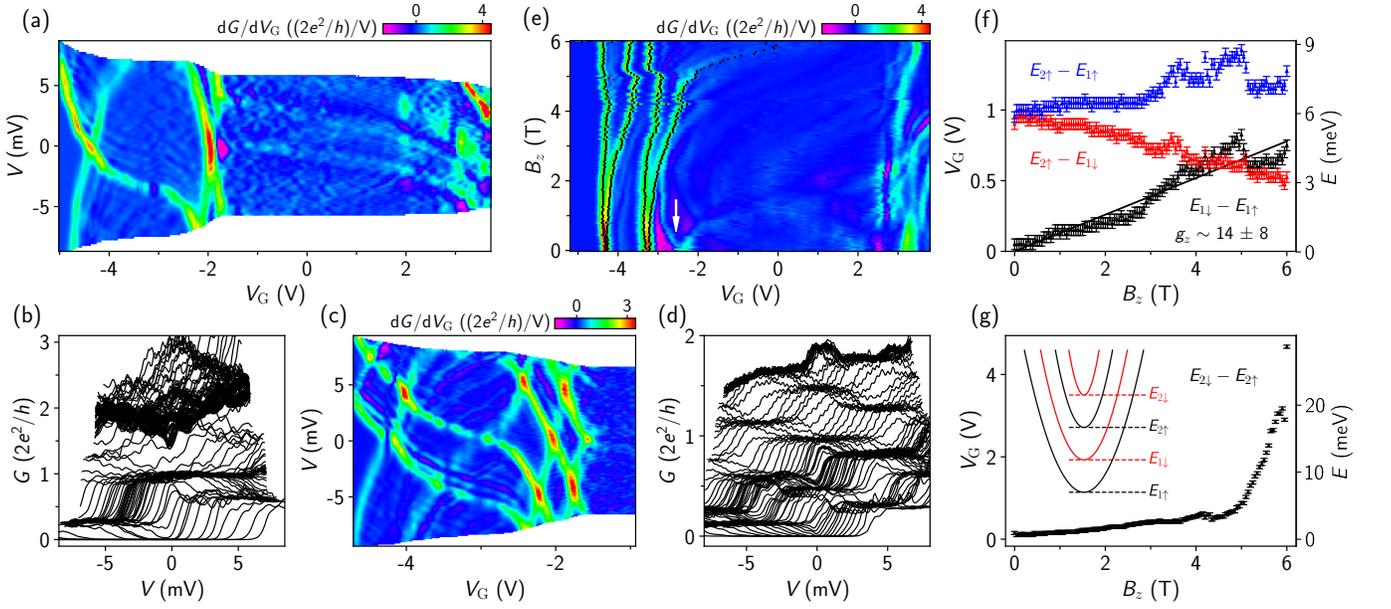

FIG. S2. Additional analysis of device B. (a) Transconductance plot of Fig. 4(a) with minor smoothing. (b) Waterfall plots of Fig. 4(a). Every other curve was plotted for clarity. The values of fractional plateaus at high biases are not exactly half, possibly due to the asymmetric circuit setup. (c) Transconductance plot of Fig. 4(c) with minor smoothing. Fabry-Pérot oscillations are also visible. (d) Waterfall plot of Fig. 4(c). Every other curve was plotted for clarity. (e) Transconductance plot of Fig. 4(e). The dots mark the positions where $G = 0.25, 0.75, 1.25$ and $1.75$ (in units of $2e^2/h$) in Fig. 4(e). The positions of the dots agree reasonably well with the transconductance peaks. The white arrow marks the quasi-bound state and its splitting in $B$. (f) Widths of the $e^2/h$ (black) and $2e^2/h$ (red) plateaus, corresponding to $E_{1\downarrow} - E_{1\uparrow}$ and $E_{2\uparrow} - E_{1\downarrow}$, respectively. The blue dots are the combined width of the two plateaus, i.e. $E_{2\uparrow} - E_{1\uparrow}$. The widths were extracted based on the spacing between the black dots in (e) and then converted to energy based on a lever arm extracted from (c). The linear fit suggests a $g$-factor of $14 \pm 8$ for the first subband. Note that the Zeeman split deviates from the linear trend, indicating a $B$ dependent $g$-factor. (g) Width of the $3e^2/h$ plateau ($E_{2\downarrow} - E_{2\uparrow}$), also exhibiting a non-linear trend of the Zeeman splitting. Inset, a sketch of the band diagram.

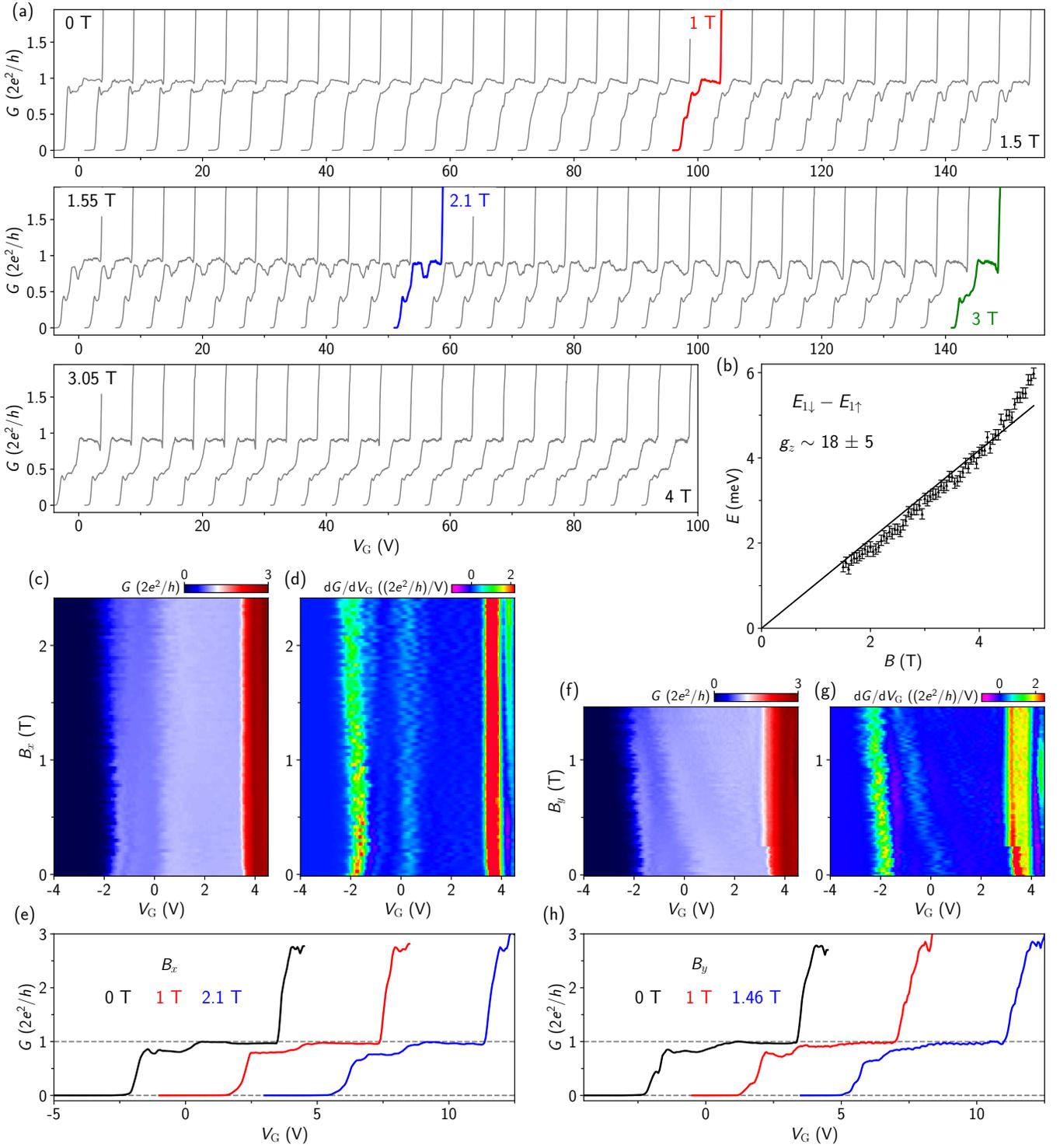

FIG. S3. Additional analysis of device C. (a) All line cuts of Fig. 5(b) for $B_z$ between 0 T and 4 T, with a horizontal offset of 5 V between neighboring curves. The step of $B_z$ is 0.05 T. (b) Width of the $e^2/h$ plateau in Fig. 5(b). Lever arm extracted from Fig. 5(i). The linear fit gives a $g$-factor of $18 \pm 5$ for $B_z$. (c) $B_x$ scan. $B_x$ is parallel to the nanowire. (d) Transconductance plot of (c). (e) Three line cuts from (c) with a horizontal offset of 4 V between neighboring curves. (f-h) Similar to (c-e) but for $B_y$. $B_y$ is in-plane and perpendicular to the nanowire. The magnet was quenched for higher $B_y$ (not shown). The absence of the $e^2/h$ plateau suggests a small $g$-factor for $B$ along these two directions.

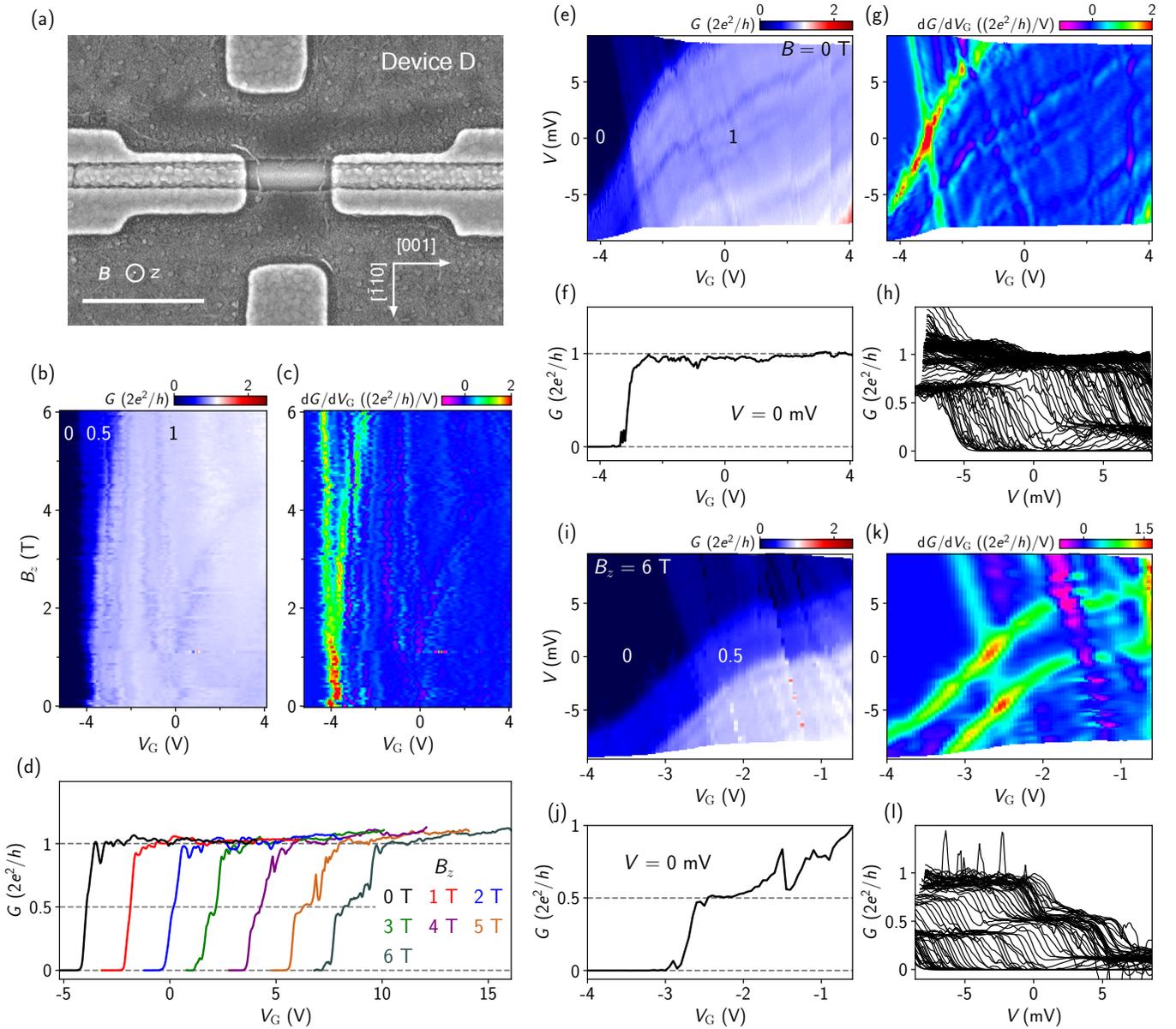

FIG. S4. Ballistic transport in device D. (a) SEM. Scale bar, 500 nm. The lower gate was fixed at 1 V while the voltage on the upper gate was scanned during the measurement. (b) $B_z$ scan of the conductance plateau. $V = 0$ mV. The labeled numbers are the plateau values in units of $2e^2/h$. $R_{\text{contact}}$ for device D is assumed to be 650 Ω. The gate starts to leak for $V_G$ beyond 4 V, higher conductance thus can not be reached through gate tuning. (c) Transconductance plot of (b) with minor smoothing. (d) Line cuts of (b) at different $B_z$'s (labeled) with a horizontal offset of 2 V between neighboring curves. (e) $G$ vs $V$ and $V_G$ at $B = 0$ T. The white diamond shape corresponds to the $2e^2/h$ plateau. Fabry-Pérot oscillations are present. (f) Zero-bias line cut of (e). (g) Transconductance of (e) with minor smoothing. (h) Waterfall plot of (e). (i) $G$ vs $V$ and $V_G$ at $B = 6$ T. The magnet was quenched when $V_G$ was scanned more positive (not shown). Conductance values (in units of $2e^2/h$) of the diamonds are labeled. (j) Zero-bias line cut of (i). The conductance drop near $V_G = -1.5$ V is due to a charge jump. (k) Transconductance of (i) with minor smoothing. The diamond size of the 0.5 plateau, $E_{1\downarrow} - E_{1\uparrow} \sim 4.6$ meV, suggesting a $g$-factor of $\sim 13$. (l) Waterfall plot of (i).

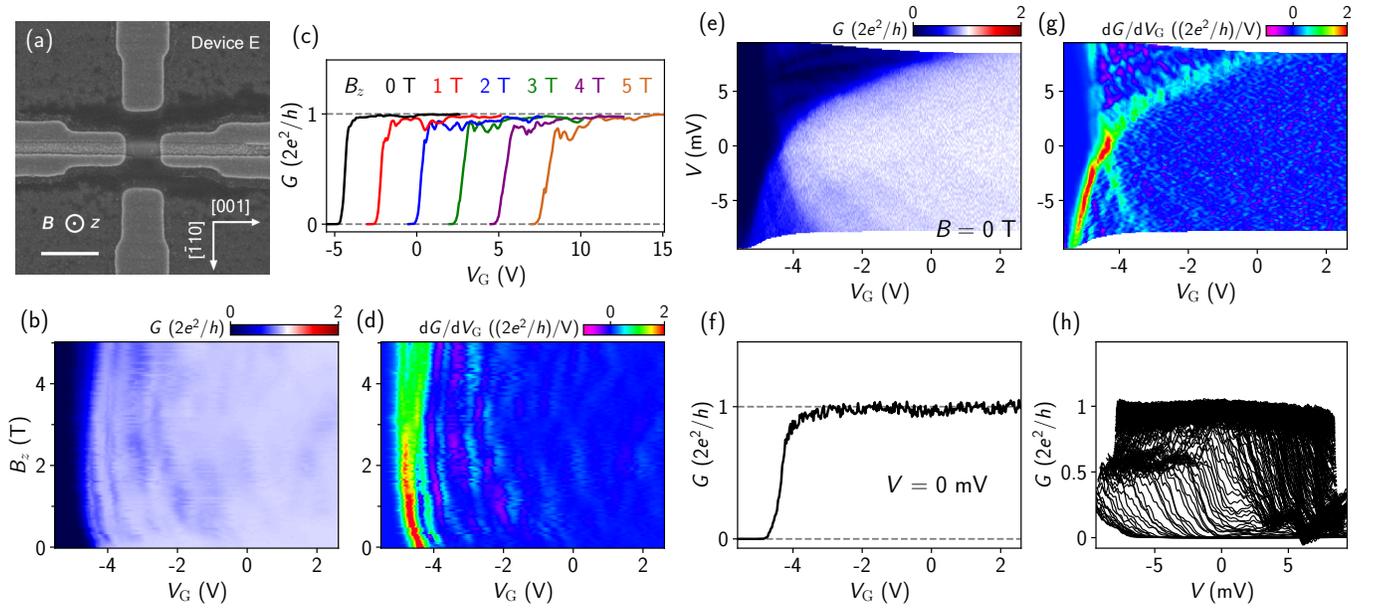

FIG. S5. Ballistic transport in device E. (a) SEM. Scale bar, 500 nm. (b) $B_z$ scan of the conductance plateau. $V = 0$ mV. $R_{\text{contact}}$ for device E is assumed to be 600 Ω. The gate starts to leak for $V_G$ beyond 2 V, thus higher conductance can not be reached. (c) Several line cuts from (b) with a horizontal offset of 2.5 V between neighboring curves. The absence of $e^2/h$ plateau at 5 T suggests a small $g$-factor. (d) Transconductance plot of (b) with minor smoothing. (e) $G$ vs $V$ and $V_G$ at $B = 0$ T. (f) Zero-bias line cut of (e). (g) Transconductance of (e) with minor smoothing. (h) Waterfall plot of (e).

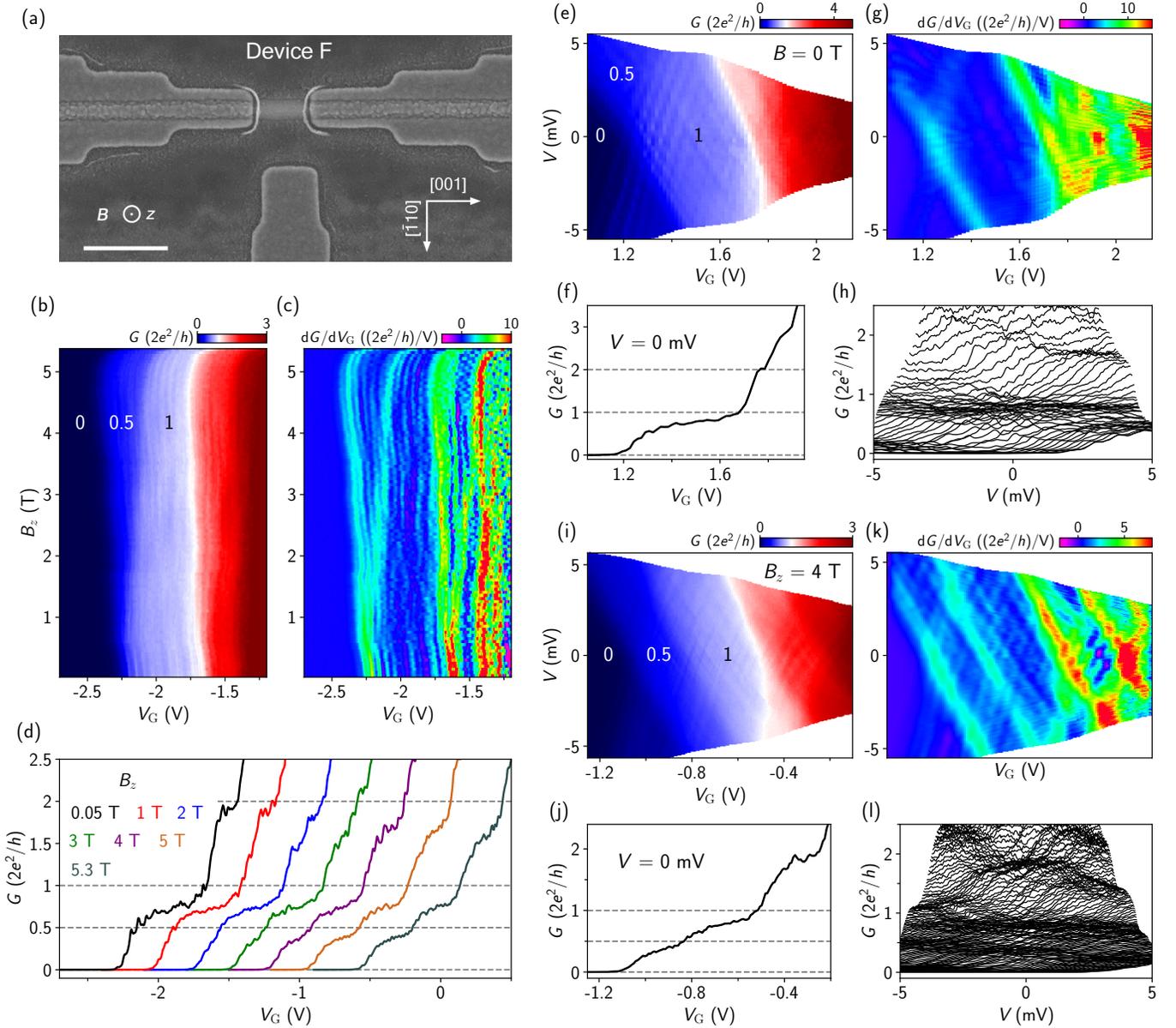

FIG. S6. Ballistic transport in device F. (a) SEM. Scale bar, 500 nm. (b) $B_z$ scan of the conductance plateau. $V = 0$ mV. The labeled numbers are the plateau values in units of $2e^2/h$. $R_{\text{contact}}$ for device F is assumed to be 1 kΩ. (c) Transconductance plot of (b) with minor smoothing. (d) Line cuts of (b) at different $B_z$'s (labeled) with a horizontal offset of 0.3 V between neighboring curves. The plateaus deviate from the quantized values, suggesting a non-unity transmission for each subband. The plateau quality of this device is worse than others. (e) $G$ vs $V$ and $V_G$ at $B = 0$ T. The diamond size of the first plateau, $E_2 - E_1 \sim 8.4$ meV. Fabry-Pérot oscillations are present. Note also the drift of $V_G$ in (e) compared to that in (b). (f) Zero-bias line cut of (e). (g) Transconductance of (e) with minor smoothing. (h) Waterfall plot of (e). (i) $G$ vs $V$ and $V_G$ at $B_z = 4$ T. Conductance values of the diamonds (in units of $2e^2/h$) are labeled. (j) Zero-bias line cut of (i). (k) Transconductance of (i) with minor smoothing. The diamond size of the 0.5 plateau, $E_{1\downarrow} - E_{1\uparrow} \sim 3.55$ meV, suggesting a $g$-factor of $\sim 15$. (l) Waterfall plot of (i).



\end{document}